# Tunneling Spectroscopy of Atomically-Thin $Al_2O_3$ Films for Tunnel Junctions

Jamie Wilt[1*], Youpin Gong[1], Ming Gong[1], Feifan Su[2], Huikai Xu[2], Ridwan Sakidja[3], Alan Elliot[1], Rongtao Lu[1], Shiping Zhao[2], Siyuan Han[1] and Judy Z. Wu[1*]

[1]*Department of Physics and Astronomy, University of Kansas, Lawrence, Kansas, 66045, USA*

[2]*Institute of Physics, Chinese Academy of Science, Beijing, China 100190*

[3]*Department of Physics, Astronomy and Materials Science, Missouri State University, Springfield, MO 65897, USA.*

Metal-Insulator-Metal tunnel junctions (MIMTJ) are common throughout the microelectronics industry. The industry standard $AlO_x$ tunnel barrier, formed through oxygen diffusion into an Al wetting layer, is plagued by internal defects and pinholes which prevent the realization of atomically-thin barriers demanded for enhanced quantum coherence. In this work, we employed *in situ* scanning tunneling spectroscopy (STS) along with molecular dynamics simulations to understand and control the growth of atomically thin $Al_2O_3$ tunnel barriers using atomic layer deposition (ALD). We found that a carefully tuned initial $H_2O$ pulse hydroxylated the Al surface and enabled the creation of an atomically-thin $Al_2O_3$ tunnel barrier with a high quality M-I interface and a significantly enhanced barrier height compared to thermal $AlO_x$. These properties, corroborated by fabricated Josephson Junctions, show that ALD $Al_2O_3$ is a dense, leak-free tunnel barrier with a low defect density which can be a key component for the next-generation of MIMTJs.

# I. INTRODUCTION

Metal-insulator-metal tunnel junctions (MIMTJs) are fundamental building blocks for microelectronics including magnetic tunnel junctions (MTJs) for spintronics and fast access nonvolatile magnetic memory, and Josephson Junctions (JJs) for particle detectors, magnetic field sensors, and qubits for quantum computation. The performance of MIMTJs depends critically on the quality of the insulating tunnel barrier[1]. Considering native oxides can naturally form on the surface of most metals, producing an atomically-thin, uniform, and pinhole-free tunnel barrier represents a major challenge in the research of MIMTJs. In Nb/Al/AlO$_x$/Nb JJs for example, an ultrathin (< 1 nm) tunnel barrier is the key to preserve phase coherence across the superconducting Nb electrodes, since the critical current ($I_c$) through the JJ exponentially decays with the barrier thickness[2]. Thermal oxidation has been the industry standard to produce AlO$_x$ tunnel barriers for JJs through *in situ* oxygen diffusion into an Al wetting layer (Fig. 1(a)). However this diffusion mediated process has difficulty achieving a uniform tunnel barrier with a well-defined thickness[3]. Despite successful commercial applications of these JJs in devices such as superconducting quantum interference devices (SQUIDs) and voltage standards, two-level defects (TLDs) in the AlO$_x$ tunnel barrier are one of the major sources of decoherence in superconducting qubits[4].

Atomic Layer Deposition (ALD) is a promising alternative for the synthesis of atomically-thin tunnel barriers for high performance MIMTJs[5, 6]. ALD is a chemical vapor process that utilizes self-limited surface reactions to grow films one atomic layer at a time (Fig. 1(b))[6, 7]. Specifically, ALD Al$_2$O$_3$ consists of a series of alternating precursor pulses of H$_2$O and trimethylaluminum (TMA) which react at the sample's surface. This process results in a fully oxidized, uniform and pinhole-free Al$_2$O$_3$ film with atomic-scale thickness control. In addition, it's reduced bulk loss tangent implies that JJs with ALD $Al_2O_3$ tunnel barriers may have a significantly reduced TLD density[8].

However, precise ALD growth and nucleation on metals remains challenging. The MIMTJ electrode and tunnel barrier deposition must be carried out *in situ* without breaking vacuum to avoid native oxides. ALD nucleation on inert metal surfaces, such as Pt and Au, can be completely frustrated for the first 30-50 cycles of alternating precursor pulses whereas for reactive metals, such as Al, even *in situ* deposited films can acquire an interfacial layer (IL) of $AlO_x$ up to ~2 nm thick[9-11]. In a previous work, *Lu et al* fabricated $Nb/Al/Al_2O_3/Nb$ JJs using *in situ* ALD of $Al_2O_3$. The presence of an IL >0.5 nm in thickness was attributed to the poor vacuum pressure (~500 mTorr) during sample transfer and pre-ALD sample heating[11-13]. This IL prevented the realization of truly atomically-thin tunnel barriers and led to poor quality JJs. Herein, we resolve these challenges by performing the sample transfer and pre-ALD heating under high-vacuum (HV) and report the first successful fabrication of atomically-thin ALD $Al_2O_3$ tunnel barriers. *In situ* scanning tunneling spectroscopy (STS) was employed to probe the growth mechanisms and physical properties of the ALD $Al_2O_3$ tunnel barriers and JJs were fabricated to illustrate the viability of ALD $Al_2O_3$ tunnel barriers for MIMTJs.

## II. EXPERIMENTAL

For samples which underwent *in situ* STS characterization, a bilayer of Nb (20 nm)/Al (7 nm) was magnetron sputtered onto a Si/Au(50 nm) substrate which was mechanically clamped to an SPM sample stage to serve as the ground contact for the Scanning Tunneling Microscope (RHK). The Au was thermally evaporated onto an updoped Si wafer with a native oxide. An *ex situ* Atomic force microscope measured it's surface roughness to be ~1.2 nm. Immediately following the Al sputtering, an aluminum oxide tunnel barrier was formed by either thermal oxidation or ALD. For the thermal oxidation samples, UHP $O_2$ was introduced to the sputtering chamber for an oxygen exposure of 1150, 1020, and 42 torr-seconds, respectively. The samples with ALD tunnel barriers were transferred to a preheated ALD chamber and then heated for 75 min or 15 min to a temperature of 200 °C - 220 °C. Following sample heating, reagents $H_2O$ and trimethylaluminium (TMA) were pulsed into the ALD chamber for 1-3 s with a purge step between pulses to deposit the ALD $Al_2O_3$ tunnel barriers.

After tunnel barrier fabrication, the samples were transferred under HV, *in situ*, to the SPM chamber which had a pressure of ~$2\times10^{-10}$ Torr. A single mechanically-cleaved Pt-Ir STM tip was used for all SPM studies. Constant height IV and dI/dV spectroscopy were taken simultaneously using the lock-in amplifier method with a voltage modulation of 100 mV at 1 kHz. The tunnel barrier height was determined by the intersection of two bisquare-method linear fits to ln(*dI/dV*). One line fit the band gap regime, and the other the conduction band. The endpoints for this linear fits were determined by eye.

The AIMD molecular dynamics simulations for the initial water activation pulse used a 2x2 supercell of FCC Al (111) under constant equilibrium volume and temperature and adopted Bohn-Oppenheimer molecular dynamics as implemented in VASP[14-16]. The NVT simulations employed the London dispersion correction using the implemented the vdW-DF functional of Langreth and Lundqvist[17] with a high plane wave energy cut-off of 450 eV to ensure a high precision. The electronic and ionic convergence criteria used were $10^{-4}$ eV and $10^{-3}$ eV respectively.

Nb-Al/ALD-$Al_2O_3$/Nb trilayers were fabricated in a homemade deposition system, which integrated UHV sputtering and ALD *in situ*[12, 18]. For comparison, traditional thermally oxidized Nb-Al/AlO$_x$/Nb trilayers were also fabricated. The Nb films were sputtered at 1.7 nm/s to minimize the formation of NbO$_x$ from trace oxygen. The sputtering chamber had a base pressure of ~$10^{-7}$ Torr or better and the sample stage was chilled-water cooled to approximately 10 °C. The bottom Nb was 150 nm, and the top Nb was 50 nm. Samples with ALD tunnel barriers were transferred *in situ* to the preheated ALD chamber, and heated for 75 min under HV. The wafer design used to investigate the quality of tunnel barriers contains 12 square junctions of four different sizes ranging from 4 μm ×4 μm to 10 μm ×10 μm. The JJ's dc current-voltage characteristics (IVC) were measured at 4.2 K in a liquid helium storage dewar.

## III. RESULTS AND DISCUSSION

### A. *In situ* Scanning Tunneling Spectroscopy and molecular dynamics simulations

ALD is a low-vacuum process that is incompatible with ultrahigh vacuum (UHV) required for both physical vapor deposition of functional electrodes and *in situ* characterization using scanning probe microscopy (SPM). To address this issue, an integrated Sputtering-ALD-SPM system was developed to allow for UHV deposition of metals, UHV SPM characterization of surfaces and interfaces, and HV ($10^{-6}$-$10^{-7}$ Torr) *in situ* sample transportation between the chambers[18]. This HV transport minimizes the metal electrode's exposure to trace gases and hence IL formation. An additional challenge to avoid IL formation is the sample heating time required to bridge the temperature difference between sputtering at 10-14 ºC and ALD at 200 °C - 220 °C. To address this challenge, the samples were inserted into a preheated ALD chamber for different times and dynamically heated to 200 °C - 220 °C under HV. Specifically, two dynamic heating times of 75 min and 15 min are presented in this work to illustrate the importance of controlling this procedure in order to achieve a clean interface between the Al and ALD $Al_2O_3$ tunnel barrier.

In Fig. 2(a), STS dI/dV spectra were taken *in situ* on Nb/Al bilayer structures (shown schematically in Fig. 2(b)) which were exposed to these two dynamic heating times. The spectrum for the 75 min heated sample (Fig. 2(a)-I) resembles that of a highly defective tunnel barrier. In fact, it has characteristics similar to the thermal $AlO_x$ tunnel barrier (discussed later in Fig. 3)[19, 20]. In contrast, the spectrum for the 15 min heated sample (Fig. 2(a)-II) closely matches the conductive spectrum measured from a calibration sample that was directly transferred to the SPM chamber after Al sputtering without going through any heating (Fig. 2(a)-II, insert). These spectra suggest that HV and short exposure between PVD and ALD are critical to minimize IL formation.

To initiate the ALD $Al_2O_3$, the Al wetting layer was exposed to a $H_2O$ pulse to hydroxylate its surface. In order to understand the kinetics of this hydroxylation process, the behavior of $H_2O$ on the Al surface was investigated using *Ab-initio* molecular dynamics (AIMD) and Climbing-Image-Nudge Elastic Band (CI-NEB) simulations. The simulation results are discussed in Fig. S1 of the Supplemental Material (SM) section. When only one $H_2O$ molecule (i.e. without $H_2O$ molecules in proximity) is present on the Al surface, $H_2O$ dissociation into $OH^-$ is thermodynamically unfavorable, as shown in Fig. S1(a)-I, II. However, when multiple $H_2O$ molecules are present on the Al (111) surface, dissociation occurs after just a few $p$s (Fig. S1(a)-III, IV). A proton transfer between nearby $H_2O$ molecules creates $OH^-$ and $H_3O^+$, followed by $H_3O^+$ dissociation into $H_2O_{ad}$ and $H^+_{ad}$. The reaction pathway of this dehydrogenation process has an overall net **exothermic** reaction with a ~0.5 eV energy barrier. The remaining transition states were verified with additional NEB simulations in Fig. S2 to have either very small or negligible energy barriers. These simulations suggest that the $H_2O$ areal density from the $H_2O$ pulse is crucial to facilitate an efficient hydroxylation reaction which will form a uniform monolayer of $OH^-$ on the Al surface. The stability of these $OH^-$ groups is also critical as dissociation into O and $H^+_{ads}$ could lead to oxygen diffusion into the Al wetting layer and IL formation. Fortunately, these $OH^-$ groups do not readily dissociate at typical ALD temperatures of ~200 °C. However, this dissociation may become a concern at significantly higher temperatures as the shown in our simulations (Fig. S3).

In order to experimentally probe this hydroxylation process, one cycle of ALD $Al_2O_3$ was performed on an Al wetting layer with an initial $H_2O$ pulse of variable duration. Figure 2(a)-III depicts a representative dI/dV spectrum for a one-cycle ALD $Al_2O_3$ tunnel barrier with an initial $H_2O$ pulse of 2 s in duration. This spectrum displays a well-defined tunnel barrier with a barrier height, $E_b$, of ~1.56 eV and indicates that an atomically-thin tunnel barrier (Fig. 2(a)-III, schematic) can be obtained using this UHV PVD-ALD approach on a clean Al wetting layer (Fig. 2(a)-II, schematic) through careful control of the ALD growth in order to minimize IL formation (Figure 2(a)-I, schematic).

Figure 2(c) reveals the one-cycle ALD $Al_2O_3$ coverage on the Al wetting layer as the initial $H_2O$ pulse duration was varied from 1-3 s. The ALD $Al_2O_3$ coverage was defined as the percentage of STS spectra, taken from random locations on the sample, which showed a sharp conduction band onset and an $E_b$ consistent with ALD samples of higher cycle number (see Fig. 3). The ALD $Al_2O_3$ surface coverage increased from ~54% at 1 s pulse duration to ~93% at 2 s duration. These experimentally observed time frames suggest that long initial $H_2O$ pulses, on the order of seconds, are required for $H_2O$ molecules, adsorbed to the Al surface, to reach a high enough areal molecular density for an efficient dissociation into $OH^-$ to occur. Interestingly, longer H$_2$O pulses were found to be detrimental to the ALD Al$_2$O$_3$ surface coverage. The remaining, non-ALD, spectra on the Al surface were either conductive or had very high noise and were unstable under the STM electric field. While the nature of these non-ALD, non-conductive spectra remains to be a topic of further investigation, we speculate that very long H$_2$O pulses may lead to H$_2$O clusters instead of monolayer formation on the Al surface. These clusters may slow down or prohibit uniform surface hydroxylation.

In addition to its paramount role in nucleation, the hydroxylation of the Al wetting layer prevents oxygen from diffusing into the Al to form an IL during the ALD process. This argument is supported by the dI/dV characteristics and $E_b$ observed for the thermal AlO$_x$ and the ALD Al$_2$O$_3$ tunnel barriers. The dI/dV spectra for a thermal AlO$_x$ tunnel barrier of ~1.3 nm, in estimated thickness[13], is shown alongside a ten-cycle ALD Al$_2$O$_3$ tunnel barrier with a comparable thickness of ~1.2 nm in Fig. 3(a). The ALD Al$_2$O$_3$ spectrum has a significantly sharper conduction band onset than the thermal AlO$_x$ spectrum, suggesting that the ALD Al$_2$O$_3$ tunnel barrier has a much more ordered and less-defective internal structure[19, 21, 22]. This improved internal structure is corroborated by the higher ALD Al$_2$O$_3$ $E_b$ shown in Fig. 3(b). Specifically, $E_b$ values of ~1.00 eV and ~1.42 eV were observed for the ALD Al$_2$O$_3$ tunnel barriers with 75 min heating and 15 min heating respectively whereas the thermal AlO$_x$ counterpart was just ~0.67 eV. Other groups have reported similar thermal AlO$_x$ $E_b$ values[13, 23]. In addition, the ALD Al$_2$O$_3$ samples with 15 min of heating had a band gap of ~2.5 eV. This high band gap is remarkable because it is comparable to the ultrathin (~1.3 nm) epitaxial Al$_2$O$_3$ band gap[24]. The ALD Al$_2$O$_3$ tunnel barrier also displayed a hard-breakdown type behavior under the STM electric field which is typical for epitaxial Al$_2$O$_3$ thin films[25]. In great contrast, the thermal AlO$_x$ tunnel barriers broke-down in a soft-breakdown manner due to defect migration within the barrier [19, 20, 25-28]. We should note that the 75 min heated samples displayed both types of breakdown, which is consistent with the thin IL found in Fig. 2(a). However the absence of soft-breakdown in the ALD $Al_2O_3$ tunnel barrier with 15 min heating can be taken as an indicator that no significant IL is present on its M-I interface.

It is particularly interesting that the ALD $Al_2O_3$ $E_b$ value was maintained as the number of ALD cycles, $N$, varied from 1 to 10 (Fig. 3(b)). This trend is particularly demonstrated in the ALD $Al_2O_3$ samples with 15 min heating (blue) and further indicates that a significant M-I IL is not present-as an IL would have disproportionately affected the samples with smaller $N$'s by lowering their $E_b$ values. For the ALD $Al_2O_3$ samples with 75 min heating (black), an IL was confirmed by the slight $E_b$ reduction of 0.11 eV as $N$ was reduced to 1 and 2 from larger values. An additional effect of this IL is demonstrated by the $E_b$ improvement as the sample heating time was reduced from 75 min (black line) to 15 min (blue line). Nevertheless, this overall ALD $Al_2O_3$ $E_b$ consistency with thickness is remarkable because it illustrates that the ALD process can produce high quality $Al_2O_3$ down to the atomically-thin limit. In contrast, the thermal AlOx $E_b$ has a significant thickness dependence in the lower nominal thickness range, although a value of 0.67 eV is maintained at 0.6-1.3 nm thickness. This $E_b$ thickness dependence is reflected by the dramatic increase in critical current density, $J_c$, observed in JJs with thermal $AlO_x$ tunnel barriers as the oxygen exposure drops below ~$10^3$ Pa-s, or ~0.4 nm in thickness[2, 13]. Furthermore, a complete tunnel barrier is not even formed in this regime as the tunneling current is dominated by pinholes.

## B. Josephson Junction characterization

To demonstrate how this ALD $Al_2O_3$ tunnel barrier performs in a demanding MIMTJ application, JJs were fabricated and their IVCs measured at 4.2 K. The IVC of a 5-cycle junction with a designed area of 100 $\mu m^2$ is shown in Fig. 4(a). This IVC has a low subgap leakage current and is highly nonlinear-as expected for Superconductor-Insulator-Superconductor (SIS) tunnel junctions. The superconducting gap voltage was $V_g \equiv 2\Delta/e \cong 2.6$ mV and did not depend on $N$. In addition, the $IR_n$ versus voltage $V$, where $R_n$ is taken to be the dynamic resistance at 5 mV, is nearly identical for JJs with different $N$; indicating good reproducibility in our junction fabrication process. These JJs are of considerably higher quality than ALD $Al_2O_3$ JJs fabricated in our previous work which had a dramatic $I_c$ suppression due to charge scatter sites in the M-I IL[12].

Recently, by measuring the dependence of JJ's critical current density on oxygen exposure, a proxy for tunnel barrier thickness $d$, Kang *et al* determined the thermal $AlO_x$ tunnel barrier $E_b$ to be ~0.64 eV [13].[1] Notice that it is very difficult to calibrate the relationship between $d$ and oxygen exposure. In contrast, due to the self-limited, layer-by-layer growth nature of ALD, the growth rate of the ALD $Al_2O_3$ tunnel barrier has been precisely calibrated as $d_{ALD} = 0.115 \pm 0.005$ nm/cycle[11]. To determine the ALD JJ $E_b$, the measured critical current density, $G_n = (R_n A)^{-1} \propto J_c$, was plotted against $d_{ALD}$ in Fig. 4(b). Because thermal and magnetic field fluctuations have a strong effect on the switching current but have essentially no effect on $R_n$, especially for JJs with small critical currents, it is much more reliable to extract $E_b$ by fitting the exponential dependence of $G_n$ versus $d_{ALD}$.

---

[1] There was a factor of 2 error in the exponential of Eq (1). Once corrected, their reported thermal $AlO_x$ $E_b$ was 0.64 eV.

$$G_n = G_0 \exp\left(-\frac{\sqrt{2m_e E_b}}{\hbar} d_{ALD}\right), \qquad (1)$$

where $m_e$ is the electron mass, $\hbar$ is the Planck constant, and $G_0$ is the specific conductance for $d_{ALD} = 0$. The tunnel barrier height determined from the best fit was $E_b = 1.10 \pm 0.06$ eV. This $E_b$ value agrees well with our STS measurements.

Ideal tunnel junctions require a uniform tunnel barrier with no microscopic pinholes. Pinholes lead to subgap leakage current and a distorted magnetic field dependence on critical current $I_c$. The magnetic field dependence of the critical current, $I_c(H)$, for a 10-cycle junction is shown in Fig. 4(c). Complete $I_c$ suppression at the first minimum and a symmetric shape was observed. This behavior is consistent with a uniform insulating tunnel barrier with negligible leakage current and pinholes.

A denser tunnel barrier should have fewer atomic-scale TLDs. TLDs have been identified as one of the major sources of decoherence for superconducting qubits, which are considered one of the strongest candidates for the implementation of scalable quantum computing[29]. It has been observed that TLDs embedded inside the oxide tunnel barrier and/or at the superconductor/oxide interface can couple strongly to Josephson qubits. These TLDs lead to splitting in the transition energy spectrum of the qubit, large fluctuations in $I_c$, and distortions in junctions' switching current distribution $P_{sw}(I)$[30-32]. Therefore, $P_{sw}(I)$ can be used as a diagnostic tool for the detection of TLDs in tunnel barriers which couple strongly to the junction. In order to reduce the effect of self-heating, a 50 µm², 10-ALD cycle junction with a very low critical current density of $J_c = 9.4$ A/cm² was selected for $P_{sw}(I)$ measurements. The critical current and shunt capacitance of this junction was $I_c = 4.757 \pm 0.003$ µA and $C \approx 2.2$ pF, respectively. Typical $P_{sw}(I)$ curves obtained at $T = 0.76$ K and 1.17 K are shown in Fig. 4(d). Using these junction parameters and a constant current sweeping rate 5 mA/s, the measured distributions agree very well with those calculated from thermal activation theory. The absence of anomalies in the measured $P_{sw}(I)$ distributions is consistent with a lack of TLDs which couple strongly to the junction in the tunnel barrier and/or at the superconductor-insulator interface.

## IV. SUMMARY AND CONCLUSIONS

In summary, an *in situ* STS study has been carried out to understand the nucleation mechanisms of ALD $Al_2O_3$ on an Al wetting layer. We have found that a well-controlled hydroxylation of the Al wetting layer, through a carefully controlled first $H_2O$ pulse, is the key to enable the creation of an atomically-thin ALD $Al_2O_3$ tunnel barrier which is of significantly higher quality than the industrial standard thermal $AlO_x$ tunnel barrier. Specifically, the ALD $Al_2O_3$ tunnel barrier has a high $E_b$ of 1.42 eV which is maintained as the barrier thickness is varied in the range of 0.12-1.2 nm. Furthermore, this ALD $Al_2O_3$ tunnel barrier has a band gap of 2.5 eV and exhibits hard electrical breakdowns similar to high-quality epitaxial $Al_2O_3$ thin films. In contrast, the thermal $AlO_x$ tunnel barrier has a low $E_b$ of ~0.67 eV only in the barrier thickness range exceeding 0.6 nm. At smaller thicknesses, enhanced soft electrical breakdown occurs and the $E_b$ decreases. Finally, the pre-ALD exposure of the Al surface in the ALD chamber, even in high vacuum, was found to be critical and must be minimized to prevent AlOx IL formation which leads to a reduced $E_b$, especially at smaller barrier thicknesses. This result demonstrates for the first time, to our knowledge, the viability of the ALD process to create an atomically-thin $Al_2O_3$ tunnel barrier which has a significantly denser, less defective internal structure than thermal $AlO_x$-as demanded for the next generation of high performance MIMTJs.

## ACKNOWLEDGEMENTS

The authors acknowledge support in part by ARO contract ARO-W911NF-16-1-0029, and NSF contracts Nos. NSF-DMR-1314861, NSF-DMR-1337737 and NSF-DMR-1508494. JW and JZW acknowledge Cindy Berrie and Jennifer Totleben for beneficial discussions on UHV SPM and assistance in synthesis of STM substrates.

**Author Contributions**

J.S.W., J.Z.W. and S.Y.H. designed the experiment. J.S.W. prepared the samples for STS and A.E. prepared JJs with the assistance of Melisa Xin. J.S.W. performed STS characterization and Y.P.G., A.E. M.G., F.F.S., H.K.X., R.T.L. carried out JJ characterization. R.S. did the simulations. All authors contributed to discussions of the results. J.S.W, S.Y.H., S.P.Z., R.S., and J.Z.W. led the effort in development of the manuscript.

**Additional information**

Supplemental Material accompanies this paper at [Supplemental Material URL]

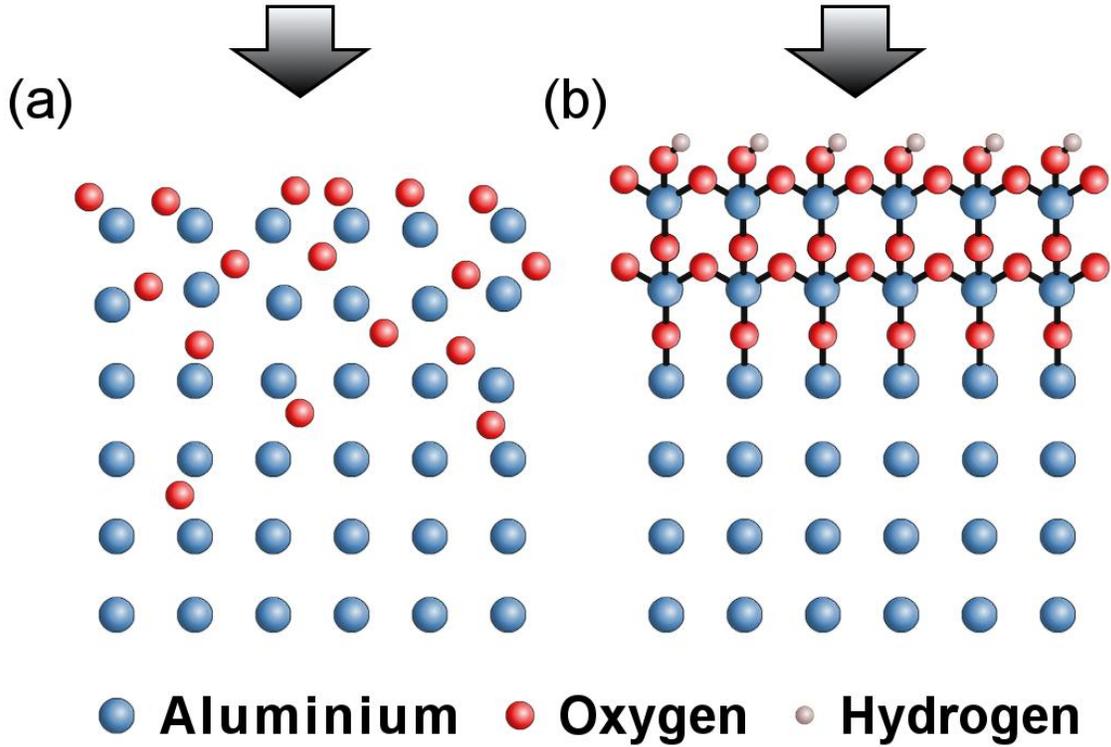

FIG. 1. Illustration which shows the structural differences between the (a) thermal $AlO_x$ tunnel barrier, formed through oxygen diffusion into the Al wetting layer, and (b) the ALD $Al_2O_3$ tunnel barrier, formed through layer-by-layer atomic layer deposition of $Al_2O_3$.

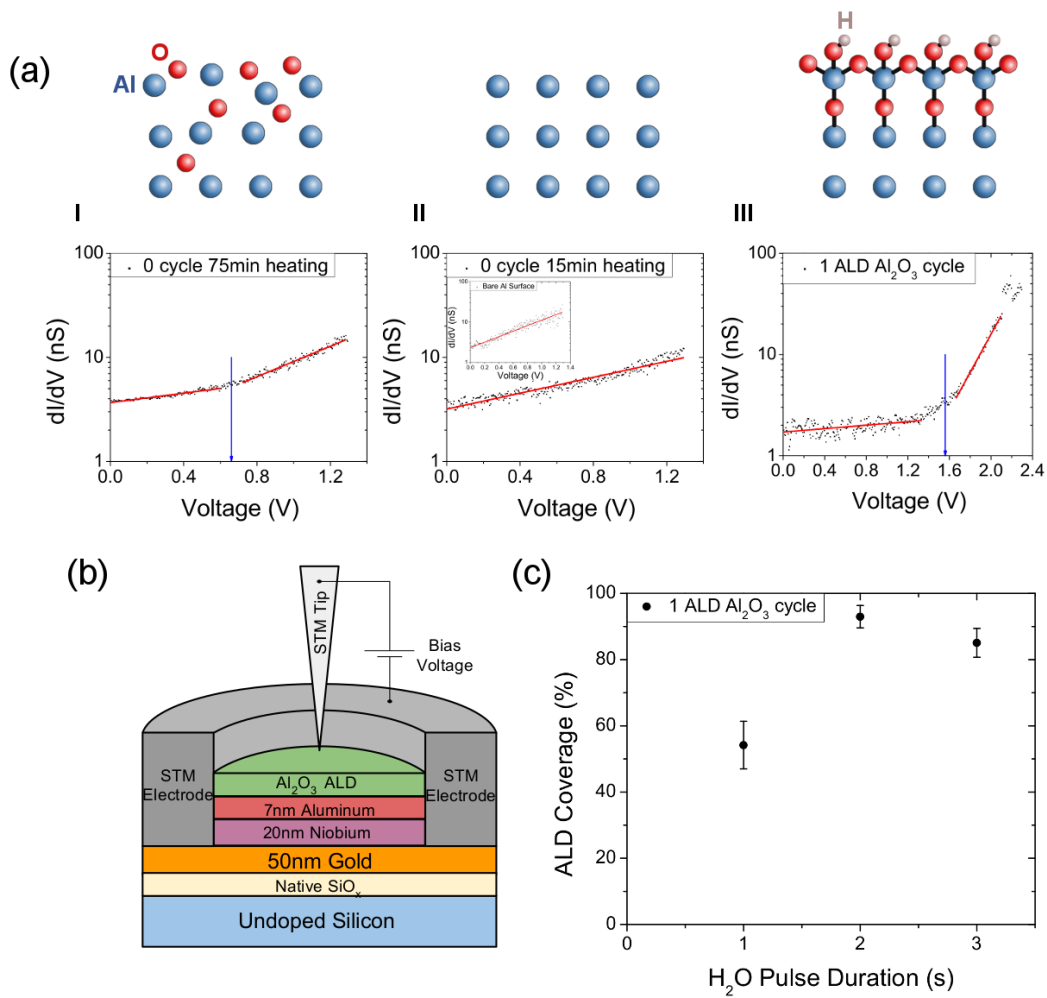

FIG. 2. Scanning tunneling spectroscopy study of the ALD $Al_2O_3$ growth on an Al wetting layer from the pre-ALD sample heating to the 1$^{st}$ ALD $Al_2O_3$ cycle (0.12 nm/cycle). (a) Exemplary dI/dV spectra are plotted for an Al sample after (I) 75 min heating in the ALD chamber, (II) after 15 min of heating, and (III) after one ALD $Al_2O_3$ cycle. The arrows (blue) depict the tunnel barrier height, calculated as the intersection of the fit lines (red). Diagrams (top) illustrate the expected surface as seen by the STM tip. The insert in (II) is the dI/dV spectrum of a sample directly transferred to the SPM chamber after Al sputtering. (b) An illustrative diagram which shows the STM sample mounting scheme. (c) The percentage of the Al surface which had a barrier height consistent with ALD $Al_2O_3$ after one ALD $Al_2O_3$ cycle with a variable initial $H_2O$ pulse duration.

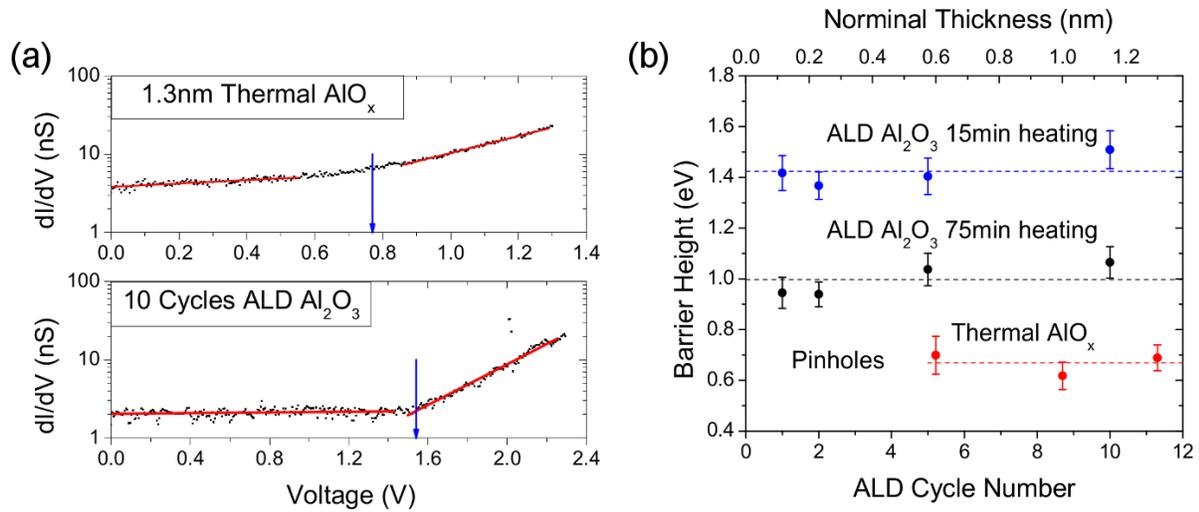

FIG. 3. A comparative scanning tunneling spectroscopy study of ALD $Al_2O_3$ vs. thermal $AlO_x$ tunnel barriers. (a) Exemplary constant height dI/dV spectra taken on a 1.3 nm thermal $AlO_x$ tunnel barrier (top) and a 10 cycle (1.2 nm) ALD $Al_2O_3$ tunnel barrier (bottom) with 15 min heating. The arrows (blue) depict the tunnel barrier height calculated as the intersection of the fit lines (red). (b) The average tunnel barrier height (dashed lines) for thermal $AlO_x$ (red) and the ALD $Al_2O_3$ (blue-15 min and black-75 min heating,) tunnel barriers plotted as function of tunnel barrier thickness respectively.

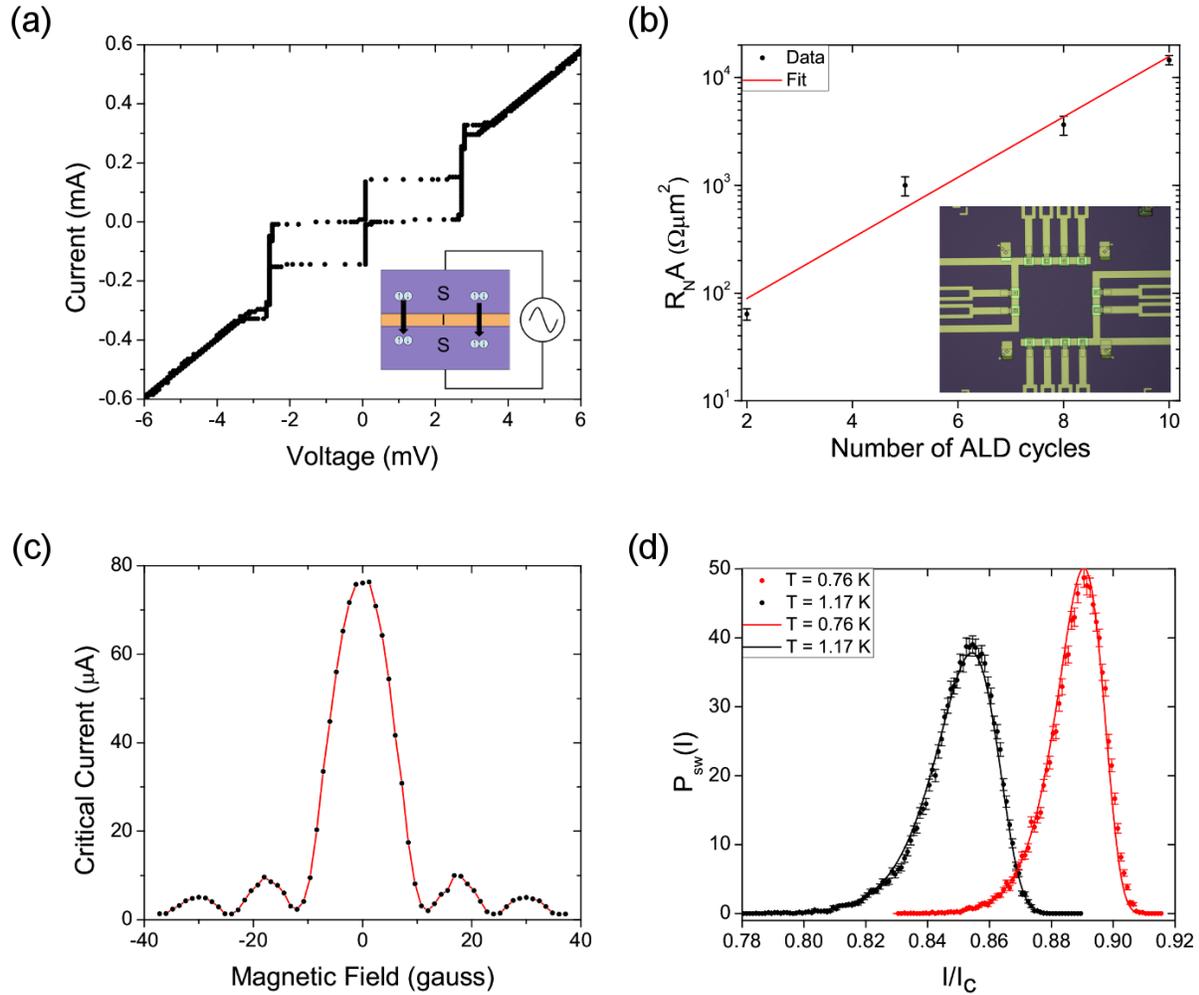

FIG. 4. Nb/Al/Al$_2$O$_3$/Nb Josephson Junctions with an ALD Al$_2$O$_3$ tunnel barrier. (a) I-V characteristics of a 5 ALD cycle 10μm x 10μm Josephson Junction which displays a very low leakage current. The insert depicts the SIS trilayer structure of the JJ with cooper pairs tunneling through the tunnel barrier. (b) The critical current density, $J_c$, as a function of ALD cycle, or equivalently thickness, which follows the expected exponential dependence (solid line). The insert shows a chip with 12 JJs with areas ranging from 25 μm$^2$ to 100 μm$^2$ fabricated using photolithography and e-beam lithography. (c) Critical current modulation by a magnetic field on a similar 5-cycle JJ processed from the same batch. (d) The measured switching current distributions (SCD) of a 10-cycle junction at T = 0.76 K and 1.17 K. The lines are calculated SCDs based on thermal activation theory.